\documentclass[aps,prl,twocolumn,groupedaddress]{revtex4-1}
\usepackage{graphicx}
\usepackage{amssymb}
\usepackage{amsmath}

\renewcommand{\Pr}{\ensuremath{\mathrm{Pr}}}
\newcommand{\Prt}{\ensuremath{\mathrm{Pr_T}}}
\newcommand{\Rk}{\ensuremath{\mathrm{R_{PK}}}}
\newcommand{\Rv}{\ensuremath{\mathrm{R_{PV}}}}
\newcommand{\Ri}{\ensuremath{\mathrm{Ri}}}
\newcommand{\Rf}{\ensuremath{\mathrm{Rf}}}
\newcommand{\Fr}{\ensuremath{\mathrm{Fr}}}
\newcommand{\Sh}{\ensuremath{\mathrm{S^\ast}}}

\newcommand{\Reb}{\ensuremath{\mathrm{Re_b}}}

\begin{document}

%\preprint{}

\title{ Asymptotic dynamics of high dynamic range stratified turbulence}

\author{ G. D. Portwood${}^{1,2}$} 
\author{ S. M. de Bruyn Kops${}^1$}
\author{ C. P. Caulfield ${}^{3,4}$}
\affiliation{${}^1$ Department of Mechanical and Industrial Engineering,
                  University of Massachusetts, Amherst, USA 01003 }
\affiliation{${}^2$ Methods and Algorithms (XCP-4),
Computational Physics Division, Los Alamos National Laboratory, Los Alamos, USA 87545}
\affiliation{${}^3$ BP Institute, University of Cambridge, Cambridge, 
  CB3 0EZ, UK}
\affiliation{${}^4$ Department of Applied Mathematics \& Theoretical
  Physics, University of Cambridge, Cambridge, 
  CB3 0WA, UK}

\date{\today}

\begin{abstract}
  Direct numerical simulations of homogeneous sheared and stably stratified
  turbulence are considered to probe the asymptotic high-dynamic range regime
  suggested by \citet{gargett84} and \citet{shih05}.  We consider
  statistically stationary configurations of the flow that span three decades
  in dynamic range defined by the separation between the Ozmidov length scale,
  $L_O=\sqrt{\epsilon/N^3}$, and the Kolmogorov length scale,
  $L_K=(\nu^3/\epsilon)^{1/4}$, up to $\Reb\equiv (L_O/L_K)^{4/3}=\epsilon/(\nu
  N^2) \sim O(1000)$, where $\epsilon$ is the mean turbulent kinetic energy
  dissipation rate, $\nu$ is the kinematic viscosity, and $N$ is the buoyancy
  frequency.  We isolate the effects of $\Reb$, particularly on irreversible
  mixing, from the effects of other flow parameters of stratified and sheared
  turbulence \citep{mater14}.  Specifically, we evaluate the influence of
  dynamic range independent of initial conditions.  We present evidence that
  the flow approaches an asymptotic state for $\Reb\gtrapprox 300$,
  characterized both by an asymptotic partitioning between the potential and
  kinetic energies and by the approach of components of the dissipation rate
  to their expected values under the assumption of isotropy. As $\Reb$
  increases above 100, there is a slight decrease in the turbulent flux
  coefficient $\Gamma=\chi/\epsilon$, where $\chi$ is the dissipation rate of
  buoyancy variance, but, for this flow,  there is no evidence of the commonly suggested $\Gamma
  \propto \Reb^{-1/2}$ dependence  when $100 \leq \Reb \leq 1000$.
\end{abstract}

\maketitle
%\pacs{}
%\keywords{}

\section{Introduction}
Sheared, stratified turbulence, energized by vertical shearing of horizontal
motions in the presence of a statically stable density distribution, arises
throughout the world's oceans and atmosphere.  Dynamical models of such flows,
in particular capturing the vertical transport of heat due to irreversible
mixing, are essential for modeling the global climate system because mixing
occurs on relatively small scales, several orders of magnitude below those
currently resolved in basin-scale models \citep{ivey08,ferrari09,gregg18b}.

A key challenge for theoreticians is to develop a robust parameterization
usable in such models for the vertical eddy diffusivity of heat $\kappa_T$
defined as
\begin{equation}
  \kappa_T \equiv \frac{B}{N^2}\equiv \frac{ \langle \frac{g}{\rho_0} w \rho\rangle }{N^2}
  , 
\   N \equiv \sqrt{-\frac{g}{\rho_0} \frac{d
    \bar{\rho}}{d z}} ,
\label{eq:kappat}
\end{equation} 
where angled brackets denote ensemble averaging, $B$ is the buoyancy flux, $N$
is the buoyancy frequency, $g$ is the constant gravitational acceleration,
$\rho_0$ is a constant reference density, $(u,v,w)$ and $\rho$ are the
fluctuating velocity vector and fluctuating density, respectively, in the
$(x,y,z)$ coordinate system, and $\bar{\rho}$ is the mean
ambient density with linear functional dependence in $z$.
Here the Boussinesq approximation has been made such that density
variations are sufficiently small %, i.e.  $\max | \rho | \ll \rho_0$, 
so that a linear equation of state is appropriate and density variations are
only significant in the buoyancy force.

 Developing dynamical models of these complex flows
 has proved 
 exceptionally difficult and controversial due not
least to their potential dependence on a wide range of parameters
\citep{mater14,ivey18}.  For simplicity, we fix $\Pr=\nu/\kappa=1$, where $\nu$ is
the kinematic viscosity and $\kappa$ is the molecular thermal diffusivity.
Even with this assumption, stratified sheared turbulence is 
influenced by at least four independent length scales, associated with
characteristic values of overall stratification, shear and both the intensity
and decay rate of turbulence \citep{mater14}.  The Ozmidov scale $L_O$, the
Corrsin (or shear) scale $L_C$, the `large-eddy' turbulent length scale $L_L$, and the
Kolmogorov microscale $L_K$ are defined here as
\begin{gather}
\begin{aligned}
   L_O &\equiv \left ({\frac{ \epsilon}{N^3}}\right )^{1/2} ;  \
  &L_C &\equiv \left ({\frac{\epsilon}{S^3}}\right )^{1/2} ; \\
   L_L &\equiv \frac{E_k^{3/2}}{\epsilon} ; \
  &L_K &\equiv \left ( \frac{\nu^3}{\epsilon} \right )^{1/4}
  , \label{eq:lengths}
\end{aligned}
\end{gather}
where $E_k$ is the averaged turbulent kinetic energy, $\epsilon$ is its
dissipation rate and $S$ is the mean constant vertical shear, i.e. $ S \equiv d
\bar{u}/d z $ where $\bar{u}$ is the mean streamwise velocity with linear
functional dependence in $z$.  These four scales are determined
by the properties of the fluctuation velocity field (i.e. the
turbulence)  relative to the background
buoyancy frequency and shear. As noted by \citet{ivey18}, it is
entirely plausible that length scales comparing the
perturbation scalar field to these background
quantities may 
be more physically relevant, 
although particularly 
in such stationary flows such as those considered here,
it is entirely plausible that the
various length scales may become closely coupled (as we investigate
further below).
%the above
%%parameterization may neglect independent length scales associated with the
%scalar which have yet to be robustly shown to exist or defined.

Although in general the scales defined in (\ref{eq:lengths}) vary in space and
time \citep[see e.g.][]{mashayek2013}, even in statistically stationary flows
where characteristic values of these four length scales can be
identified, parameterization of key properties of the flow could depend on at least
three independent nondimensional parameters determined
from these scales. Choices for these parameters are a characteristic Richardson
number, $\Ri$,  a turbulent Froude number $\Fr$ and the activity parameter
$\Reb$ (sometimes called the buoyancy Reynolds number and formally distinct from a related integral-scale quantity; see
 \citet{portwood16} for further discussion) defined as
 \begin{equation}
\begin{gathered}
  \Ri \equiv \frac{N^2}{S^2} \equiv \left ( \frac{L_C}{L_O} \right )^{4/3} ,\\ %\label{eq:ridef} \\
  \Fr \equiv  \frac{\epsilon}{N E_k}  \equiv \left (\frac{L_O}{L_L} \right ) ^{2/3} , %\label{eq:frtdef}
  \Reb \equiv  \frac{\epsilon}{\nu N^2} \equiv \left (\frac{L_O}{L_K} \right )^{4/3} , %\label{eq:gndef}
\end{gathered}
  \label{eq:nd_defs}
\end{equation}
although alternative parameters can be defined, such as a `shear number'
$\Sh\equiv (L_L/L_C)^{2/3}\equiv S E_k/\epsilon \equiv 1/(\Fr \Ri^{1/2})$
and others \citep{mater14}.% and a Reynolds number $\Re\equiv (L_L/L_K)^{4/3}\equiv
%E_k^2/(\nu \epsilon) \equiv \Reb/\Fr^2$. 
%GDP: We dont use Re anywhere, so this is cut for space

In terms of these parameters, the central challenge regarding $\kappa_T$
becomes that of determining the functional dependence of
$\kappa_T(\Ri,\Fr,\Reb)$. In a profoundly influential paper, Osborn
\citep{osborn80} argued from consideration of the turbulent kinetic energy
equation that in stationary flows it is reasonable to suppose that the
buoyancy flux $B$, or equivalently, for the stationary flows considered here, the dissipation rate of buoyancy variance
$\chi$ defined as
\begin{equation}  
  \chi \equiv \kappa \left \langle \frac{g^2}{\rho_0^2 N^2}{\left |  \nabla \rho \right |^2 }\right \rangle 
  , \label{eq:chi}
\end{equation}
can be linearly related to the dissipation rate $\epsilon$ through a `turbulent
flux coefficient'  or `mixing efficiency' $\Gamma$, i.e.  $\chi=\Gamma
\epsilon$.   Osborn hypothesized an upper bound for $\Gamma \leq 0.2$ on
semi-empirical grounds, although  in practice, $\Gamma$ is implemented as a
constant saturating the upper bound\citep{gregg18b}. 
The introduction of $\Gamma$ transforms the fundamental issue of modeling
$\kappa_T$ into identifying the functional dependence of
$\Gamma(\Ri,\Fr,\Reb)$, since $\kappa_T= \nu \Gamma \Reb$ provided the flow is
statistically stationary.

Many models have been presented for the functional dependence of $\Gamma$
based on observations, experiments and numerical simulations in a variety of
different flows, with much recent activity
\citep[e.g.][]{ivey91,barry01,shih05,ivey08,
  maffioli16b,venaille17,salehipour16,mashayek17, zhou17,gregg18b,monismith18}
though with little sign of consensus, due principally to three fundamental
issues.  Issue {\it I} is that disentangling transient and/or reversible
processes from the irreversible processes crucial for quantifying and
parameterizing mixing is extremely challenging.  Issue {\it II} is that, even
for flows where the assumption of stationarity is appropriate, stratification
in a gravitational field and vertical shear both break isotropy, and so the
extent to which anisotropy is important is difficult to determine,
particularly when there is a relatively small dynamic range between $L_K$ and
$\min (L_C, L_O)$ (as encountered by \citet{gargett84}).
The condition $L_C < L_O$ is associated with growing and stationary flow
configurations, as necessarily $\Ri < 1$, suggesting that shear instabilities have
to be sufficiently vigorous to sustain turbulence.  This has the consequence
that $L_C \gg L_K$ is a typical condition for high dynamic range, i.e. $\Ri
\Reb \gg 1$ by the relations defined in \eqref{eq:nd_defs} and as suggested by \citet{itsweire93}.  
Finally, issue {\it III} is that although in principle $\Reb$, $\Ri$ and $\Fr$
are all independent parameters, there is emerging evidence
\citep[e.g.][]{lucas17} that in many flows the parameters become correlated,
and so an apparent dependency of $\kappa_T$ or $\Gamma$ on one parameter is
actually associated with variation in another parameter.

To address all three issues, we consider the model flow of {\em stationary
  homogeneous sheared and stratified turbulence} (S-HSST)%, which has been
%applied to the study of high Reynolds numbers turbulence
\citep[e.g.][]{holt92,itsweire93,jacobitz97,shih05}.  In S-HSST, the production
of turbulent kinetic energy by uniform mean vertical shear exactly balances the
dissipation rates of kinetic and potential energy by molecular motion,
addressing issue {\it I} by ensuring energetic stationarity by design.  When
numerically simulated, the turbulence fills the flow domain provided that $\Reb$
is sufficiently large \citep{stillinger83}.  More generally, stratified
turbulence has been shown to exhibit different dynamics depending on the value
of $\Reb$ \citep{shih05,hebert06a,bartello13,debk15,debk19}.  \citet{gargett84}
observed in limited two-point statistics from ocean data that
Kolmogorov-Oboukhov-Corrsin (KOC) scaling may be observed in stratified
turbulence when $\Reb \sim O(10^{3})$ and higher.  \citet{debk15} observes that
KOC scaling is generally not observed in DNS up to $\Reb=220$, though some statistics
may be consistent with such scalings. 
An important open
question is thus how stratified turbulence behaves when $\Reb > O(100)$, a
parameter space accessible by modern simulations of S-HSST with 
sufficiently small $\Fr$ for the stratification to have a first-order effect on
the turbulent dynamics. However, to address issue {\it II} there is the further
constraint that $\Ri \Reb \gg 1$.

In fact, in S-HSST $\Ri$ and $\Fr$ cease to be free, but rather must adjust to
ensure statistical stationarity.  It has been empirically observed in
simulations that the various dynamic length scales `tune' so that
$\Fr\approx0.5$ and $\Ri\approx 0.16$, and equivalently $\Sh \approx 5$.  An
immediate consequence of the emergence of an apparently $\Reb-$invariant
stationary value of $\Ri$ in S-HSST is that issue {\it II} can indeed be
addressed at sufficiently large $\Reb$.

Perhaps more importantly, issue {\it III}  is also  addressed, as the
dependence of flow properties on $\Reb$ can be probed independently of the other
parameters.  Specifically, although the adjustment of $\Fr \approx 0.5$ and
$\Ri \approx 0.16$ occurs at $\Reb \sim O(10)$, other dynamical scalings have
been observed to change as $\Reb$ increases beyond $\Reb \sim O(100)$
 \citep{shih05,debk15}.  The objective of this study is thus to investigate how the
key properties of the turbulence, in particular the energy partitioning, mixing
and small-scale anisotropy depend on variations of $\Reb$.  A specific question
to investigate is whether the postulated power law $\Gamma \propto
\Reb^{-1/2}$ \citep{shih05,ivey08,monismith18} for sufficiently large $\Reb$
occurs in this flow in which dependence on other flow
parameters can be completely eliminated.  

\section{Simulations} 
The incompressible Navier-Stokes equations are considered, subject to the
nonhydrostatic Boussinesq approximation and coupled with equations for
continuity and buoyancy transport.  A turbulent decomposition is performed
relative to a mean buoyancy gradient and a mean streamwise velocity gradient,
both in the vertical direction
\citep[see also ][]{rogallo81, holt92, jacobitz97, shih00}.  We implement the
system with the same Fourier pseudo-spectral scheme described by
\citep{almalkie12a,debk15}, except that an additional shear term is handled by
an integration factor \citep{brucker06,chung12,sekimoto16}. 

%In the continuous S-HSST configuration, the fluctuating
%quantities relative to the time-constant planar means are homogeneous.  The
%time-discrete problem is not inherently homogeneous \citep[see][for more
%  details]{chung12} but is solved using a mimetic numerical method that
%preserves the homogeneity inherent in the continuous equations.

Stationarity is induced by fixing a value of $\nu$, choosing a target turbulent
kinetic energy $E_t$ then adjusting the Richardson number via 
$g$ \citep[c.f.][]{taylor16} using a mass-spring-damper control system: 
\begin{equation}
  c_0 S \Ri'(t)+2\alpha \omega \tilde{E}_k'(t) + \omega^2 (\tilde{E}_k(t)-1)=0
\end{equation}
where the prime notation denotes a temporal derivative, $\tilde{E}_k(t)\equiv
E_k/E_t$ is the normalized turbulent kinetic energy, $\omega$ is
the characteristic
frequency of oscillation and $\alpha$ is a dimensionless damping factor.  The
control system has been derived by assuming that the
kinetic energy follows a second order linear system \citep[e.g.][]{rao11},
and then by applying the first-order approximation $\tilde{E}_k'(t)\approx c_0 S (\Ri(t)-\Ri_c)$
such that $\tilde{E}_k''(t) \approx c_0 S \Ri'(t)$.  The parameter $c_0\approx -1$
is supported by \citet{jacobitz97}, the characteristic frequency $\omega$ is determined by the 
mean shear, and a damping coefficient $\alpha=1.5$ was found to work well.

Crucially, following this procedure, $\Ri \approx 0.16$ emerges without
presupposition for all our cases, as shown in the table \ref{tb:table1} along
with other emergent parameters; flow statistics are averaged over a period of
$St \approx 100$ unless noted otherwise.  
\begin{table}
  \begin{ruledtabular}
    \begin{tabular}{r c c c c c  }
      Case      &  $\Reb$  & \Ri & \Fr & $N_x$\\ \hline
      SHSST-R1  &  36  & 0.163 & 0.46 & 1024  \\
      R2        &  48  & 0.159 & 0.47 & 1280  \\
      R3        &  59  & 0.162 & 0.48 & 1536  \\
      R4        &  81  & 0.154 & 0.50 & 1792  \\
      R5        &  110 & 0.155 & 0.52 & 2048  \\
      R6        &  160 & 0.157 & 0.48 & 3072  \\
      R7        &  240 & 0.156 & 0.48 & 4096  \\
      R8        &  390 & 0.146 & 0.46 & 6144  \\
      R9        &  550 & 0.163 & 0.45 & 8192  \\
      R10       &  900 & 0.152 & 0.42 & 9600 
    \end{tabular}
  \end{ruledtabular}
  \caption{
  Simulation parameters.  $N_x$ is the number of grid points in the
  $x$-direction and the grid spacing is isotropic.
  \label{tb:table1}}
\end{table}
Furthermore, the dissipation rate $\epsilon$ also `tunes' such that $\Fr
\approx 0.5$.  As the dissipation rate is an emergent quantity, the smallest
length scales are resolved by adjusting the resolution such that $k_{max}L_K
\approx 2$, where $k_{max}$ is the largest Fourier domain wavenumber.
We also found it necessary to use a relatively
large domain  with $L_x/L_y=2$, $L_x/L_z=4$ and
$L_x/L_L\approx40$, where $L_x$, $L_y$, and $L_z$ are the dimensions of the
domain, in order to support the anisotropic large scales of the flow.

%This methodology allows a wide range of $\Reb$ to be
%accessed.  We note that the case SHSST-R1 corresponds to the lowest $\Reb$
%for which stationarity could be maintained, near the threshold for active
%turbulence proposed by \citet{gibson80} and others
%\citep[e.g.][]{stillinger83,itsweire93}.  

\section{Results and discussion}
\subsection{Energetics}
\begin{figure}
   \includegraphics{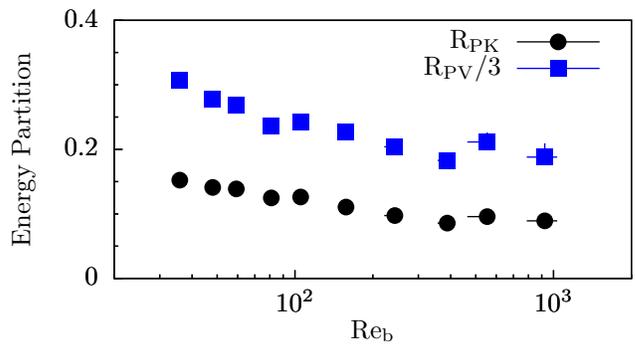}
   \caption{ Energy partitions as a function of $\Reb$.  Bars indicate upper and
   lower quartile measurements of instantaneous quantities. 
\label{fig:fig1} }
 \end{figure}

We stress that, although $E_k \approx E_t$ is enforced, the parameters $\Ri$, $\Fr$ and $\Reb$ emerge from the
simulations, as do the structure of the turbulence and the scalar.
We consider the ratio of the potential energy to kinetic
energy, $\Rk\equiv E_p/E_k$, and the ratio of the potential energy to the
kinetic energy of vertical motion, $\Rv\equiv E_p/E_v$ where the energies are
defined as:
\begin{equation*}
  E_v \equiv \frac{1}{2}\langle {w^2} \rangle \;\text{and}\; 
  E_p \equiv \frac{1}{2}\left<\frac{g^2/\rho_0^2}{N^2}    {\rho^2} \right>\text{ .}
\end{equation*}
Energy-partitioning is a critical component to mixing models wherein Reynolds
number, or $\Reb$, dependence is often omitted and the mixing is assumed to be a
function of $\Ri$ \citep{osborn72,schumann95,pouquet18}.  Furthermore, in
`strongly' stratified turbulent flow, 
 \citet{billant01} %page 1649 
suggest that there should be approximate equipartition between
potential and kinetic energy, i.e. $\Rk\approx1$, an assumption also used by \citet{lindborg06a}.
Figure \ref{fig:fig1} illustrates that this
basic assumption of equipartition is not appropriate. 
Perhaps more interesting, the energy ratios decrease by a factor of two
relative to the lowest $\Reb$ case until $\Reb \approx 300$, after which the
energy ratios appear to remain constant with anisotropy of velocity variance
explaining the different behaviors of $\Rk$ and $\Rv$. 
%, an effect
%which naturally reduces as $\Reb$ increases.
%GDP: We find the above statement to not be true for the velocity variance
%

\subsection{Small-scale anisotropy}
The observation of anisotropy in the energy partitions of the previous section
is particularly important, due not least to the inevitable requirement to
estimate energetic dissipation rates in circumstances where each component of
the rate-of-deformation tensor is not available.  The existence of anisotropy
at dissipative length scales would require dynamical estimates of $\kappa_T$ to
account for anisotropy.  Therefore, it is necessary to evaluate common
dissipation surrogate models to assess their applicability and to
characterize small-scale anisotropy in these flows.  Here, we use single
component surrogate models which rely on a single derivative and the isotropy
assumption:
\begin{subequations}
   \begin{align}
     \tilde\epsilon_{ij} &= 
     \begin{cases}
       15 \nu \left \langle \big(\frac{\partial u_i}{\partial x_j}\big)^2 \right \rangle & \text{if $i=j$} \\
       15/2 \nu \left  \langle  \big(\frac{\partial u_i}{\partial x_j}\big)^2 \right \rangle & \text{if $i\neq j$}
     \end{cases} \\
     \tilde\chi_j &= 3 \kappa \left \langle\frac{g^2}{\rho_0^2 N^2} \left ( \frac{\partial \rho}{\partial x_j}\right )^2 \right \rangle \text{ .}
   \end{align}\label{eq:diss}
\end{subequations}

In flows that are inherently anisotropic due to mean shear or mean flux, it is
widely assumed that small-scale isotropy is a reasonable assumption when the
scale separation is large \citep[e.g.][]{gargett84}.  However, evidence
suggests that isotropy assumptions can be very inaccurate in stably stratified
flows  at finite $\Reb$ \citep{itsweire93,hebert06a,debk15,debk19}.
%\citet{debk15} noted that the isotropic
%assumption is reasonable in homogeneous stratified turbulence provided
%$\Reb\gtrapprox200$.
%, yet postulated that an even higher $\Reb$ threshold
%may be appropriate.  
%
\begin{figure*}[htp]
  \includegraphics{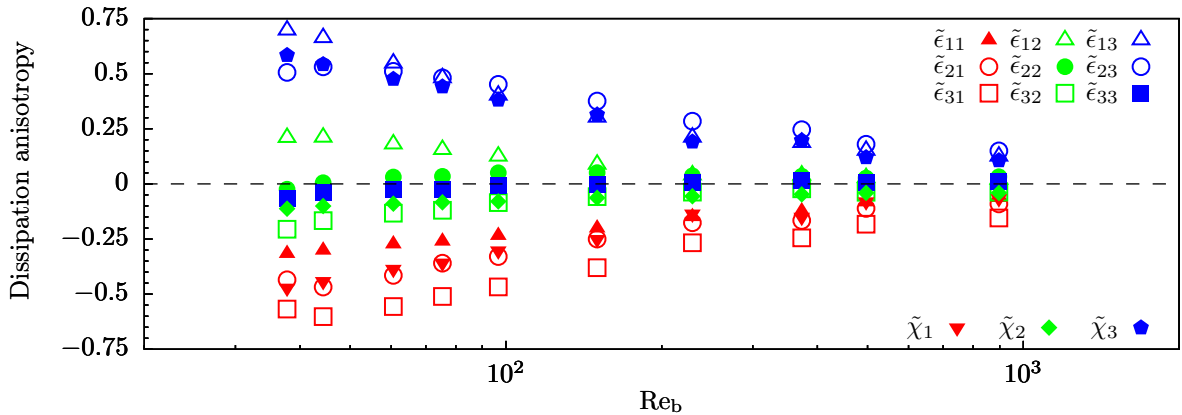}
  \caption{ Instantaneous measurements of small-scale anisotropy.  The dashed line represents perfect isotropy at small scales.  \label{fig:fig3}}
\end{figure*}

The validity of \eqref{eq:diss} is tested with the help of figure
\ref{fig:fig3} in which is plotted the relative error $(\tilde\epsilon_{ij} -
\epsilon)/\epsilon$ of the various single-component dissipation estimates.
As with energy partition, there is an apparently asymptotic regime for $\Reb
\gtrapprox 300$ in which the assumption of isotropy applied to dissipation
rate becomes valid within approximately 15\% error. 
Nevertheless, there is evidence of
small-scale anisotropy as expected from the analysis in \citet{durbin91}, which shows that
dissipation-range isotropy should only exist when $\Sh \ll 1$; in these
simulated flows $\Sh \approx 5$.  However,
we observe that $\tilde\epsilon_{22}$, $\tilde\epsilon_{33}$, and $\tilde\chi_{2}$ 
yield good estimates of the dissipation rates even in cases in
which the small scales are strongly anisotropic. 

\subsection{Mixing coefficient}
\begin{figure}[htp]
  \includegraphics{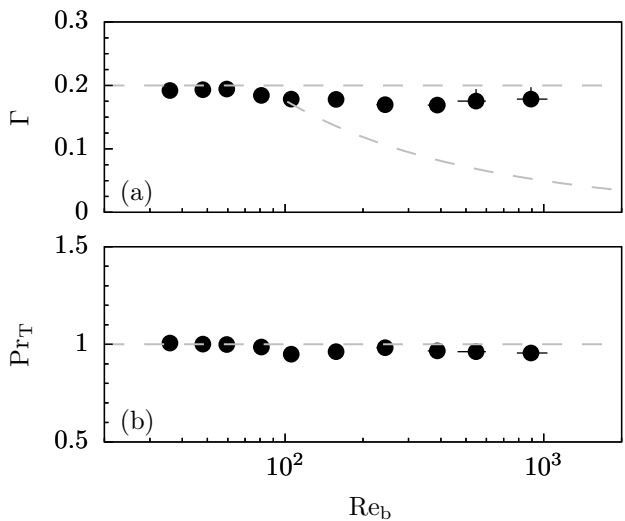}
   \caption{ a: Variation of turbulent flux coefficient $\Gamma$ with $\Reb$.  The upper bound proposed by Osborn is indicated by the gray dashed line where $\Gamma \approx 0.2$, and the $\Reb^{ -1/2 }$-based parameterization suggested in \citet{shih05} is plotted for their `energetic' regime of $\Reb > 100$.  b: Variation of turbulent Prandtl number $\Prt$ with $\Reb$. 
\label{fig:fig2}
   } 
\end{figure}

We plot the turbulent mixing coefficient $\Gamma \equiv \chi/\epsilon$ in
figure \ref{fig:fig2}a.
Whereas there is some modest decrease with increasing $\Reb$ from the
peak $\Gamma$ of approximately 0.19, remarkably close to Osborn's suggested
upper bound, there is no evidence of the commonly-suggested
$\Reb^{-1/2}$ scaling \citep[e.g.][]{shih05,ivey08,monismith18}. A decrease in mixing
efficiency with respect to $\Reb$ is observed until $\Reb \approx
200$, above which $\Gamma \approx 0.17$. 
There is no evidence of the `energetic' regime of \citet{shih05} for $\Reb >
100$, suggesting that transient multi-parameter effects may well be relevant in the
evolution of their flows such that $\Reb$ is not always an independent parameter (see issue {\it III} as described in \S 1). 
There is evidence
that the upper bound proposed by Osborn is a useful estimate, at least in flows
where the underlying assumption of stationarity is well-justified, as
such flows 
naturally adjust to $\Ri \approx 0.16$ and $\Fr  \approx 0.5$.

Alternative descriptions of the mixing behavior are 
instructive.  In the asymptotically high $\Reb$ regime, the buoyancy variance dissipation rate
adjusts such that $\chi \approx E_p N$ as evidenced by figures \ref{fig:fig1} and \ref{fig:fig2}. Although this is a natural scaling, it
is of interest that the $O(1)$ constant is actually extremely close to $1$.  A
second instructive description of mixing is the turbulent Prandtl number
$\Prt= \kappa_M/\kappa_T$, where $\kappa_M$ is the eddy diffusivity of
momentum:
\begin{equation}
  \kappa_M \equiv \frac{\langle u w \rangle}{S} =
  \frac{P}{S^2}, \label{eq:kappam}
\end{equation}
and $P$ is the production rate of turbulent kinetic energy.  Therefore, 
\begin{equation}
  \Prt= %\left ( 
  \frac{N^2}{S^2} \frac{P}{B}\equiv \frac{\Ri}{\Rf} \approx
  \frac{\Ri (1+\Gamma)}{\Gamma} , \label{eq:prt}
\end{equation}
where $\Rf$ is the `flux Richardson number', and the relationships $P\approx
B+\epsilon=\chi+\epsilon$ in a stationary flow has been used.  $\Prt$ is
plotted in \ref{fig:fig2}b, and proves to be close to one for all
values of $\Reb$. This is perhaps unsurprising, but makes clear that the
turbulent processes that mix heat and momentum in these flows are highly coupled,
and in particular that the stratification is not sufficiently `strong' to
modify the turbulent processes greatly, but rather that the
irreversible conversion of
kinetic into potential energy occurs in a 
balanced, equilibrated way.

Using 
mixing length arguments, \citet{odier09} defined:
\begin{equation}
  L_\rho \equiv \left ( \frac{B}{N^2 S} \right )^{1/2} ;
  \
  L_m    \equiv \left ( \frac{P}{S^3} \right )^{1/2} ;\label{eq:odier}
\end{equation}
such that $L^2_m/L^2_\rho=Pr_T$. Therefore, since for
our flows $Pr_T\approx1$ and $P \approx (1+ \Gamma) \epsilon$,
it is apparent that the density length scale at the heart
of the model presented by \citet{ivey18} is coupled
to the Corrsin scale by $L_\rho/L_C=\sqrt{1+\Gamma}$, and the
key parameter $D\equiv \chi/\epsilon=1/\Gamma$ 
has nearly fixed value. 

\section{Conclusions}
We have used a canonical controlled flow to evaluate the isolated effects
of high dynamic range in sheared stratified turbulent flow.  The energy
partitioning varies non-trivially where $\Reb \lessapprox 300$ above which an
apparently asymptotic regime is entered, consistent with
\citep{gargett84,debk15}.  The flow retains measurable anisotropy at
dissipation scales, as suggested by the analyses of \citet{durbin91}, at all
values of $\Reb$ we have considered.  Nevertheless, some single-component
surrogates exist which accurately estimate dissipation rates via isotropy
assumptions even for smaller $\Reb$.

The results presented here seem to indicate that the effects of the large
dynamic-range regimes explored by Gargett et al.\ are strongly influenced by
asymptotic scalar density dynamics, rather than by the velocity field
independently.  Nevertheless, the measured mixing is a much
weaker function of $\Reb$ compared to some proposed scalings \citep{shih05},
with the turbulent Prandtl number $\Prt \approx 1$ and the turbulent mixing
coefficient $\Gamma$ near the classical bound of $0.2$ as suggested by Osborn,
although decreasing slightly for $\Reb \lessapprox 300$.  The application of
these results to higher Prandtl numbers merits further study, where dynamic
range arguments indicate transition at lower $\Reb$ as $\Pr$ is increased. 

We stress that the flow we have considered has, by design, controlled
dependence on all other parameters.  $\Ri \approx 0.16$ and $\Fr \approx 0.5$
naturally emerge to ensure stationarity.  Therefore, we conjecture that
observed apparent variation of mixing properties with $\Reb$
\citep{monismith18} can be explained by breaking one or the other of these
constraints, i.e. the variation is due to either transient effects, well-known
to lead to strong variation in mixing properties
\citep[e.g][]{salehipour16h,mashayek2017t}, or `hidden' and perhaps correlated
variation with other parameters \citep[e.g][]{zhou17}.  Additionally, there
could also be as yet unquantified strong dependence on initial or boundary
conditions, whereas the flow we have considered is isolated from such
effects. To develop robust mixing parameterizations, it is necessary to
develop appropriate models to capture the effects of such conditions, informed
by and generalizing from such controlled, idealized flows as considered here.

%\clearpage
\begin{acknowledgments}
This work was funded by the U.S.\ Office of Naval Research via grant
N00014-15-1-2248.  High performance computing resources were provided through
the U.S.\ Department of Defense High Performance Computing Modernization
Program by the Army Engineer Research and Development Center,  the Army
Research Laboratory and the Navy DSRC under Frontier Project
FP-CFD-FY14-007.
The research activity of C.P.C. is supported by EPSRC Programme Grant
EP/K034529/1 entitled `Mathematical Underpinnings of Stratified
Turbulence'.
Valuable comments from anonymous reviewers have substantially improved
the clarity of the discussion.
\end{acknowledgments}

%\bibliography{bib}
%

\end{document}